\documentclass[conference,a4paper]{IEEEtran}
\IEEEoverridecommandlockouts
\usepackage[sort]{cite}
\usepackage{amsmath,amssymb,amsfonts}
\usepackage{algorithmic}
\usepackage{graphicx}
\usepackage{textcomp}
\usepackage{xcolor}
\usepackage{booktabs}
\usepackage{array}
\usepackage{multirow}
\usepackage{xurl}
\usepackage{float}
\usepackage{listings}
\usepackage[colorlinks=true, citecolor=blue!50!black, linkcolor=blue!50!black, urlcolor=blue!50!black, breaklinks=true]{hyperref}
\usepackage{placeins}
\usepackage{flafter}
\usepackage[a4paper,left=13mm,right=13mm,top=19mm,bottom=43mm]{geometry}
\def\BibTeX{{\rm B\kern-.05em{\sc i\kern-.025em b}\kern-.08em
    T\kern-.1667em\lower.7ex\hbox{E}\kern-.125emX}}

\newif\ifanonymous
\anonymousfalse

\begin{document}

\title{BugGen 2.0: Rethinking the Roles of LLMs for Scalable RTL Bug Synthesis}

\ifanonymous
\author{Anonymous Authors}
\else
\author{
\IEEEauthorblockN{Minh Luu, Michael Quinn, Aakash Tyagi, David Houngninou, Jiang Hu}
\IEEEauthorblockA{\textit{Texas A\&M University} \\
College Station, Texas, USA \\
\{minh.luu,m.d.quinn,tyagi,davidkebo,jianghu\}@tamu.edu}
}
\fi

\maketitle

\begin{abstract}

AI-assisted RTL localization and debugging need ground-truth bug datasets with known locations and observable failures, while testbench grading uses injected bugs to measure detection capability. Creating such datasets at scale remains a major EDA challenge. BugGen 1.0 shows that Large Language Models (LLMs) can generate context-aware RTL mutations. Still, it relies on the model for both choosing where to mutate and generating the edit, which increases inference costs and can lead to invalid choices. BugGen 2.0 changes this division of labor. Deterministic algorithms handle structural analysis, target selection, and validation, while the LLM focuses on localized semantic rewriting. Across four OpenTitan IP blocks and FabScalar with three LLMs, BugGen 2.0 reaches every requested quota of compilable bug scenarios that cause at least two test failures, with 14--16 of 16 configured mutation types represented. In completed same-model comparisons, it reduces token use by 6.8\(\times\)--15.4\(\times\) relative to BugGen 1.0 and 1.5\(\times\)--7.9\(\times\) relative to Firefly. BugGen 2.0 also expands the dataset without simulating every new bug. After finding accepted scenarios with observable failures, it generates alternative mutations at the same locations, simulates only a subset, and estimates acceptance for the rest. This skips 63.6\% of AES and 33.9\% of FabScalar variant simulations. Full simulation finds that 77.7\% of AES variants and 95.2\% of FabScalar variants are accepted, with a 1.09\% estimation error on the directly comparable FabScalar set. These results show that focusing LLMs on semantic rewriting lowers inference and simulation cost while improving consistency across models and designs.

\end{abstract}


\section{Introduction}

Growing design complexity makes verification and debugging a dominant part of integrated circuit (IC) development, with design functional verification (DFV) alone accounting for more than half of the overall design effort~\cite{wilson2024}. Bug datasets can support both machine learning (ML)-based localization and debugging~\cite{vcdiag,wang2026simfree}, and systematic grading of verification environments, but obtaining useful ground truth at scale remains difficult. Manual fault injection does not scale, while predefined RTL mutation testing frameworks~\cite{mantra} can generate many faults but capture limited semantic and architectural context. More broadly, preparing open datasets remains a major obstacle to applying ML in EDA~\cite{nsf2023mleda}. The practical need is therefore for a scalable way to generate varied, context-aware RTL bugs and evaluate them in industrial-style verification flows.

We refer to BugGen~\cite{buggen} as BugGen 1.0 throughout this paper. Its LLM-centered workflow partitions RTL modules, then uses three agents to select a region, choose a mutation type and target, and generate the edit. A shared cache records successful outcomes, but this history influences later decisions only through in-context reasoning rather than enforced constraints. As a result, incompatible choices can consume several model calls before compilation rejects them, and rollback may lead the agents back to similar unsuccessful decisions. Since prior outcomes and prompt constraints remain advisory, the LLM can repeat earlier mistakes or generate edits that violate those constraints.

BugGen 2.0 explicitly handles decisions with deterministic mechanisms rather than relying on the LLM. \textbf{Type-aware filtering} removes incompatible targets before generation using regex rules and abstract syntax tree (AST)-derived types. \textbf{Stratified line sampling} spreads selections across the RTL file, while \textbf{compile learning} uses previous outcomes to steer mutation type selection. \textbf{Batched generation} requests several localized edits in one model call. Finally, once a bug scenario is accepted, \textbf{variant augmentation} generates alternative mutations at its locations, simulates only a subset, and estimates the outcomes of the rest.

This paper makes three major contributions that, to our knowledge, are new to RTL bug synthesis. First, it introduces a hybrid workflow that assigns structural analysis, target selection, and validation to deterministic algorithms while reserving the LLM for localized rewriting. Second, it introduces variant augmentation, which reduces simulation by evaluating only a subset of generated alternatives and estimating the rest. Third, it characterizes model dependence by evaluating efficiency and consistency across different LLM backends. Across five designs and three LLMs, BugGen 2.0 reaches every requested quota of accepted bug scenarios and covers 14--16 of the 16 configured mutation types. On AES and FabScalar, it also uses substantially fewer tokens than LLM-centered baselines.

\section{Related Work}

\subsection{Traditional Hardware Mutation Testing}

Traditional RTL mutation testing frameworks~\cite{mantra} apply predefined syntactic mutations primarily to assess verification quality, with undetected mutants revealing gaps in the testbench. Although scalable, these approaches capture limited semantic and architectural context. Prior research on mutation testing in software has also explored reducing execution costs through historical operator performance~\cite{petrovic2021mutation} and predictive detection models~\cite{jain2023predictive}. BugGen 2.0 applies these ideas to RTL bug synthesis, using localized RTL context to generate diverse mutations that produce observable failures for debugging and learning. It therefore combines context-aware RTL mutation with history-guided, budgeted evaluation.

\subsection{LLM-Based RTL Mutation}

Recent approaches use LLMs to move beyond predefined mutation operators by generating edits from RTL context. Firefly~\cite{firefly} prompts an LLM with complete RTL files and relies on the model to select both mutation locations and types. BugGen 1.0~\cite{buggen} adds a closed-loop multi-agent workflow with cached history, but still delegates structural decomposition and target selection to LLM reasoning. Related work, such as RTLFixer~\cite{rtlfixer}, instead uses EDA tool feedback to guide LLM-based RTL repair rather than to generate bugs. Thus, existing LLM-based bug synthesis systems continue to rely on the model for both target selection and edit generation. BugGen 2.0 separates these roles, using explicit structural analysis and selection mechanisms to reduce inference costs while retaining a broad representation of mutation types.

\section{Introducing BugGen 2.0}

\begin{figure}[!t]
\centering
\includegraphics[width=0.80\columnwidth]{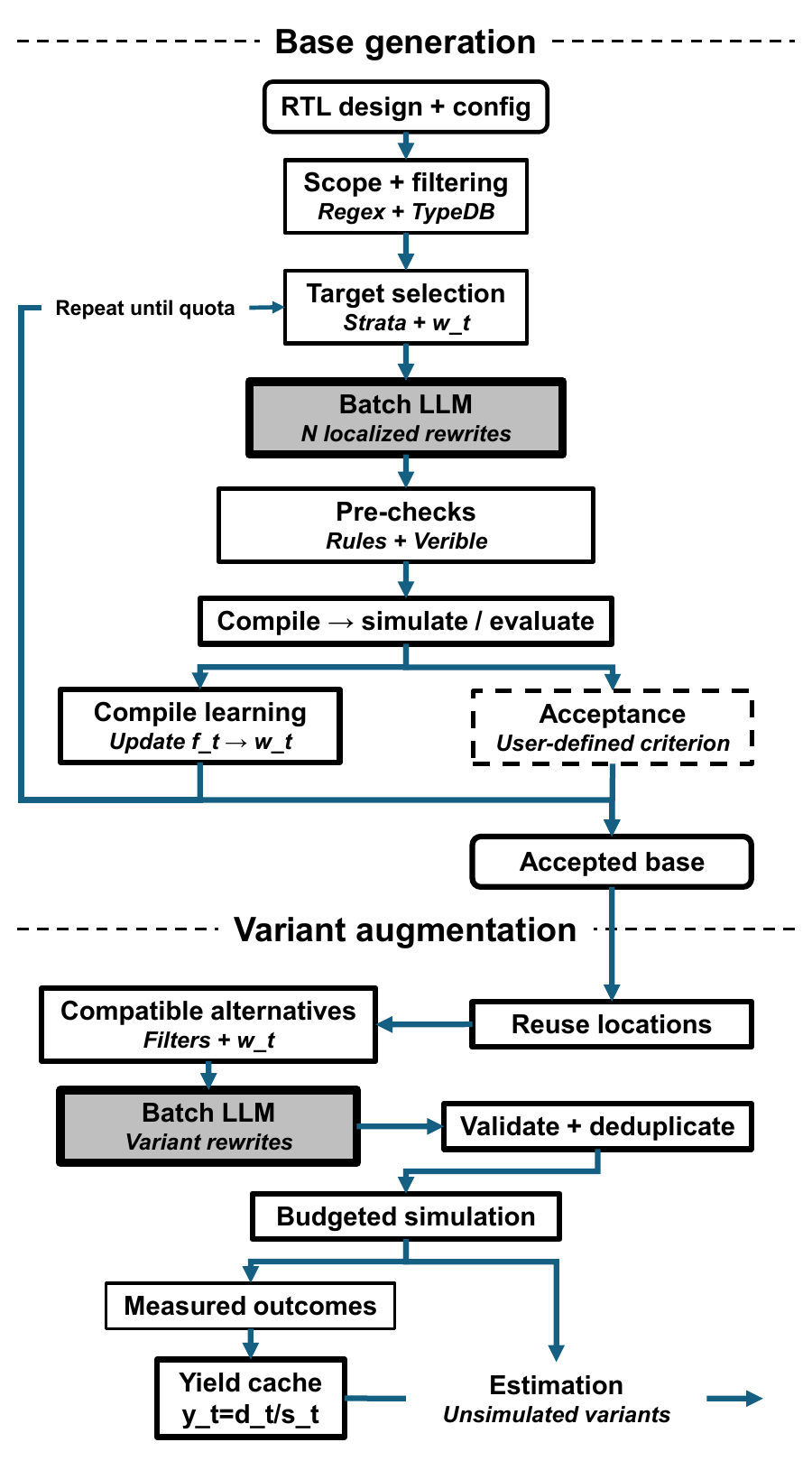}
\caption{BugGen 2.0 pipelines.}
\label{fig:buggen2_architecture}
\end{figure}

Fig.~\ref{fig:buggen2_architecture} shows the base generation and variant augmentation pipelines. BugGen 2.0 first determines which RTL locations and mutation types are structurally compatible, and then asks the LLM to generate only localized edits. Generated candidates pass through syntax, compilation, and simulation checks. This ordering removes unsuitable choices before they consume more expensive inference and verification resources.

A \emph{mutation} is one RTL code edit whose injection location is known by construction, while a \emph{bug scenario} is a modified RTL instance containing $N\geq1$ mutations. The main pipeline produces \emph{base scenarios}. Once a base scenario is accepted, augmentation can generate alternative mutations at the same locations, producing \emph{variants}. For example, after a scenario containing a mutation of type $t$ at line $\ell$ causes observable test failures, augmentation can keep $\ell$ fixed and generate other compatible edits there. A scenario is \emph{accepted} if it compiles and satisfies the user-defined criterion for its generation stage. In our experiments, base generation requires at least two failing tests, and variant augmentation requires at least one.

\subsection{Scope and Filtering}

Before generation, BugGen 2.0 extends the BugGen 1.0 rulebook with mutation-specific regex rules and a signal-type database, TypeDB, extracted from the Verible AST~\cite{verible} using a technique similar to VCDiag~\cite{vcdiag}. These checks map each RTL line to compatible mutation types and remove known incompatibilities, including mismatched user-defined types, enums, or structs, before invoking the LLM or compiler. A signal swap, for instance, is rejected when the candidate signals have incompatible types or widths.

\subsection{Target Selection}

After filtering, BugGen 2.0 samples among the compatible mutation types using a weight updated from prior outcomes, as described in Section~\ref{sec:compile_learning}. To avoid repeatedly targeting the same part of a file, the system divides each file into $R$ approximately equal strata, or contiguous line ranges. For each of the $N$ mutations, it selects the least-used eligible stratum not already used in the scenario, breaking ties uniformly at random, then selects the least-used compatible line within that stratum.

\subsection{Batch Generation and Pre-checks}

For each selected mutation type and line pair $(t,\ell)$, BugGen 2.0 extracts a local context window around $\ell$. All selected windows are packed into one prompt, so that a three-mutation scenario can request all three edits in one LLM call. The LLM returns the edits as a JSON array, similar to Firefly~\cite{firefly}. If an entry is missing or malformed, the valid edits are retained, and only the failed slot is regenerated using another compatible mutation type.

Each returned edit is checked against the original RTL and its compatibility rules. Valid edits are then applied one at a time to a clean copy and parsed with Verible~\cite{verible} before full compilation. An earlier edit can change the source text around a later target, so a later slot is omitted if its target can no longer be found. The resulting scenario may therefore contain fewer than the requested $N$ mutations.

\subsection{Compile Learning}\label{sec:compile_learning}

Compile learning uses prior outcomes to guide future mutation type selection. Intuitively, a mutation type that repeatedly produces invalid RTL becomes less likely to be selected again within the same module. Each type has a nonnegative penalty score $f_t$: failures increase the score, accepted scenarios decrease it, and larger scores reduce future selection.

A scenario may contain several mutations, so the system cannot always determine which mutation caused a failure. It therefore applies the same scenario-level update to every mutation type used in the scenario. This avoids relying on noisy, tool-specific compiler logs, but does not assign causal blame to a particular mutation.

For the currently eligible mutation type set $\mathcal{T}_e$, the sampling probability of type $t$ is
\begin{equation}
\pi(t)=\frac{w_t}{\sum_{j\in\mathcal{T}_e}w_j},
\qquad
w_t=\alpha^{\min(f_t,\,f_{\max})}\rho_t,
\end{equation}
where $\alpha$ suppresses types with prior failures, $\rho_t$ penalizes reuse within the current scenario, and $f_{\max}$ caps the penalty. By default, $\alpha=0.75$, so each additional failure multiplies the base weight by 0.75. We set $\rho_t=0.7$ if type $t$ has already been selected in the current scenario and $1$ otherwise, with $f_{\max}=20$. Scores begin at zero and reset at module boundaries so that feedback remains local to each module.

After observing scenario outcome $s$, the score is updated as
\begin{equation}\label{eq:update_rule}
f_t \leftarrow \max\!\left(0,\,f_t+\delta_s\right),
\end{equation}
where $\delta_s$ is $+\beta$, $-\gamma$, or $+\delta_u$ for a failure, accepted scenario, or compiled but unaccepted scenario, respectively. We use $(\beta,\gamma,\delta_u)=(1,1,0)$. Failures include LLM no-operation outputs, Verible syntax failures, and compiler failures. A compiled but unaccepted scenario leaves the score unchanged because the result may reflect testbench observability rather than mutation quality.

\subsection{Variant Augmentation for Data Expansion}
\label{sec:augmentation}

Variant augmentation expands the dataset by reusing locations from accepted base scenarios. Since these locations come from base scenarios with observable failures, BugGen 2.0 explores different compatible mutations there rather than searching for new targets. Specifically, if a base scenario modifies lines $\ell_1$ and $\ell_2$, augmentation keeps those locations fixed and replaces the original mutation types with alternatives.

For each accepted base, augmentation generates up to $v_{\max}$ variants. At each location, the original type is excluded, while the same regex, TypeDB, and no-operation checks determine compatible replacements. Types are sampled using compile learning weights with a reuse penalty $\rho_t$ for types already selected in the scenario. All variants are generated in one LLM call, applied to clean RTL copies, and checked for compatibility and Verible validity. Invalid or duplicate variants are discarded.

Simulating every variant would be expensive, so BugGen 2.0 evaluates only a subset and estimates acceptance for the rest. For retained variant set $V_s$ of base $s$ and nominal simulation fraction $b$, the per-base budget is
\begin{equation}
\label{eq:variant_budget}
B_s =
\begin{cases}
0, & |V_s|=0, \\
\max\!\left(1,\mathrm{round}(|V_s|b)\right), & |V_s|>0.
\end{cases}
\end{equation}

At least one simulation ensures that every base contributes measured evidence, so the total fraction can exceed $b$. Consider 125 bases, each producing four or five variants, yielding 600 in total. With $b=0.2$, the per-base minimum gives 125 simulations instead of 120. The remaining question is which variants to simulate. A type with weight $w_t$ near 1 has accumulated little penalty and is already favored by the generation policy, while a type near 0 has accumulated repeated penalties and is strongly suppressed. BugGen 2.0, therefore, spends more of the limited simulation budget on types between these extremes using $u_t=4w_t(1-w_t)$. The score peaks at $u_t=1$ when $w_t=0.5$ and decreases toward 0 at either extreme. At $w_t=0.1$ or $0.9$, for instance, $u_t=0.36$. Since $w_t$ is a ranking weight, not a probability, this is a mid-weight selection heuristic and does not represent uncertainty estimation.

Unsimulated variants are estimated from prior simulated outcomes. For each mutation type, BugGen 2.0 records the observed acceptance rate $ y_t = d_t/s_t$, where $d_t$ and $s_t$ are the counts of accepted and simulated variants, respectively. An unsimulated variant receives the mean rate of its mutation types, or the design's base acceptance rate if no history exists. A variant may contain multiple mutations, so $y_t$ reflects association with acceptance, not the causal effect of one type. Measured and estimated outcomes are reported separately, and the cache persists across runs within a design.

\section{Experimental Setup}

We evaluate BugGen 2.0 on four OpenTitan~\cite{opentitan} IP blocks (AES, OTP Controller, ROM Controller, and USB) and the FabScalar~\cite{fabscalar} processor core. OpenTitan uses SystemVerilog UVM testbenches with Synopsys VCS~\cite{synopsys_vcs}. FabScalar uses Verilog, a C++ instruction-set simulator as its reference model, and Cadence Xcelium~\cite{cadence_xcelium}. Table~\ref{tab:design_specs} summarizes these flows. Each experiment uses a predefined accepted-scenario quota. A scenario contains at most $N=3$ simultaneous mutations, and base acceptance requires compilation followed by at least $\tau=2$ failing tests.

\begin{table}[!ht]
\caption{Evaluated designs and verification flows.}
\label{tab:design_specs}
\centering
\small
\setlength{\tabcolsep}{4pt}
\renewcommand{\arraystretch}{1.05}
\begin{tabular}{@{}lccc@{}}
\toprule
\textbf{Design} & \textbf{RTL Lines} & \textbf{Verification} & \textbf{Simulator} \\
\midrule
AES       & 17,198 & UVM      & VCS \\
OTPCtrl   & 12,747 & UVM      & VCS \\
RomCtrl   &  2,691 & UVM      & VCS \\
USB       & 15,942 & UVM      & VCS \\
FabScalar & 40,575 & ISS Ref. & Xcelium \\
\bottomrule
\end{tabular}
\end{table}

For comparison, we rerun BugGen 1.0 and Firefly~\cite{firefly} on the same RTL and LLM backends. We adapt only their interfaces to the verification flows, preserving their prompting, target selection, and mutation generation procedures. AES also includes a commercial RTL mutation testing tool as a traditional reference. All LLM-based frameworks use the same 16 mutation types in Table~\ref{tab:mutation_types}.

Experiments use Google Gemma 4 31B (G4)~\cite{gemma4}, OpenAI gpt-oss-120b (OSS)~\cite{gptoss}, and DeepSeek v4 Flash (D4F)~\cite{deepseekv4}. BugGen 2.0 uses a 32,768-token context window, an 8,192-token output cap, disabled reasoning, provider-default sampling parameters, and no model-specific tuning. Token counts include prompts and completions from every request, including retries and unsuccessful outputs. We report token-reduction factors only for the same model.

The systems use different stopping rules. BugGen 1.0 and BugGen 2.0 run toward accepted scenario quotas, whereas Firefly receives one attempt per configured scenario. As a result, attempts should not be compared as a common budget across all rows. Tokens per accepted scenario (kTok/A) instead normalizes inference volume by accepted output.

\begin{table}[!ht]
\caption{Mutation types available for all LLM-based frameworks.}
\label{tab:mutation_types}
\centering
\scriptsize
\setlength{\tabcolsep}{2pt}
\renewcommand{\arraystretch}{1.05}
\begin{tabular}{@{}l@{\hspace{7pt}}l@{}}
\toprule
\multicolumn{2}{c}{\textbf{Mutation Types}} \\
\midrule
\texttt{missing\_assignment}
& \texttt{signal\_swap} \\
\texttt{bitwise\_corruption}
& \texttt{stuck\_at\_fault} \\
\texttt{control\_flow\_bug}
& \texttt{default\_value\_change} \\
\texttt{wrong\_assignment}
& \texttt{pipeline\_hazard} \\
\texttt{incorrect\_data\_size}
& \texttt{checker\_corruption} \\
\texttt{fsm\_transition}
& \texttt{register\_address\_corruption} \\
\texttt{adjacent\_field\_swap}
& \texttt{handshake\_corruption} \\
\texttt{loop\_modification}
& \texttt{masking\_corruption} \\
\bottomrule
\end{tabular}
\end{table}

\section{Results}\label{sec:results}

The results answer three questions: how much inference and wall-clock time the hybrid workflow uses, how its components affect validation and line-position spread, and how much additional data variant augmentation adds under a limited simulation budget.

\subsection{End-to-End Performance}

Table~\ref{tab:baseline_comparison} reports AES and FabScalar results. \emph{A/Att.} is accepted scenarios over top-level attempts, where one attempt may contain several internal generation retries. \emph{C} is the fraction of generated candidates that pass Verible and compilation. \emph{kTok/A} is the total number of LLM tokens per accepted scenario, including retries and unsuccessful outputs. \emph{T} counts mutation types represented in accepted scenarios, and \emph{S} counts post-compilation simulation evaluations. Each evaluation may execute several tests.

\begin{table}[!t]
\caption{End-to-end performance on AES and FabScalar.}
\label{tab:baseline_comparison}
\centering
\small
\setlength{\tabcolsep}{2.2pt}
\renewcommand{\arraystretch}{1.00}
\begin{tabular}{@{}llccccc@{}}
\toprule
\textbf{System} & \textbf{Model}
& \textbf{A/Att.} & \textbf{C}
& \textbf{kTok/A} & \textbf{T} & \textbf{S} \\
\midrule

\multicolumn{7}{@{}l}{\textbf{AES}} \\

\multirow{3}{*}{Firefly}
& G4  & 56/140  & 45\% & 16.2 & 8  & 63 \\
& OSS & 94/140  & 74\% & 10.5 & 11 & 103 \\
& D4F & 56/140  & 42\% & 41.6 & 13 & 59 \\

\addlinespace[2pt]
\multirow{3}{*}{BG 1.0}
& G4$^\dagger$  & 62/100 & 50\% & 89.4 & 10 & 95 \\
& OSS$^\dagger$ & 62/85  & 62\% & 80.8 & 11 & 101 \\
& D4F           & 140/203 & 85\% & 81.2 & 11 & 333 \\

\addlinespace[2pt]
\multirow{3}{*}{BG 2.0}
& G4  & 140/147 & 80\% & 4.62 & 15 & 163 \\
& OSS & 140/153 & 71\% & 7.15 & 15 & 163 \\
& D4F & 140/148 & 78\% & 5.26 & 15 & 173 \\

\midrule
\multicolumn{7}{@{}l}{\textbf{FabScalar}$^\ddagger$} \\

\multirow{3}{*}{Firefly}
& G4  & 339/340 & 100\% & 12.1 & 10 & 340 \\
& OSS & 333/340 & 98\% & 10.6 & 12 & 333 \\
& D4F & 324/340 & 96\% & 14.6 & 15 & 325 \\

\addlinespace[2pt]
\multirow{3}{*}{BG 1.0}
& G4  & 360/503 & 100\% & 80.6 & 9  & 859 \\
& OSS & 360/361 & 96\%  & 41.9 & 13 & 387 \\
& D4F & 360/371 & 96\%  & 45.2 & 7  & 435 \\

\addlinespace[2pt]
\multirow{3}{*}{BG 2.0}
& G4  & 360/390 & 93\% & 8.31 & 14 & 481 \\
& OSS & 360/364 & 86\% & 6.17 & 14 & 381 \\
& D4F & 360/361 & 90\% & 4.17 & 14 & 394 \\

\bottomrule
\end{tabular}

\vspace{1.5pt}
\parbox{\columnwidth}{%
\footnotesize
$^\dagger$ BugGen 1.0 AES G4 and OSS did not reach the 140-scenario quota within 48 hours.
$^\ddagger$ Firefly targets 17 unique RTL files (17 \(\times\) 20 = 340 scenarios), while BugGen has 18 target entries (18 \(\times\) 20 = 360 scenarios) because \texttt{RegRead.v} appears in two groups. Thus, the FabScalar workloads differ slightly.
}

\end{table}

BugGen 2.0's efficiency, as shown in Table~\ref{tab:baseline_comparison}, is less sensitive to the choice of LLM. Across both designs, its token cost remains within 4.17--8.31 kTok/A, while Firefly varies more strongly on AES and BugGen 1.0 incurs much higher cost when repeated retries fail to make progress. In same-model comparisons, BugGen 2.0 reduces token use by 15.4\(\times\) on AES with D4F and 6.8\(\times\)--10.9\(\times\) on FabScalar, while the reduction relative to Firefly is 1.5\(\times\)--7.9\(\times\).

\begin{figure*}[!t]
\centering
\begin{minipage}[t]{0.32\textwidth}
    \centering
    \includegraphics[width=\linewidth]
        {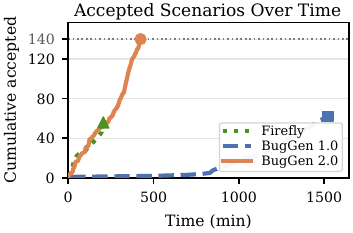}
    \vspace{-1mm}
    {\footnotesize\textbf{(a)} AES cumulative acceptance (G4)}
\end{minipage}
\hfill
\begin{minipage}[t]{0.32\textwidth}
    \centering
    \includegraphics[width=\linewidth]
        {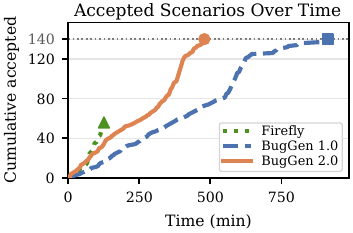}
    \vspace{-1mm}
    {\footnotesize\textbf{(b)} AES cumulative acceptance (D4F)}
\end{minipage}
\hfill
\begin{minipage}[t]{0.32\textwidth}
    \centering
    \includegraphics[width=\linewidth]
        {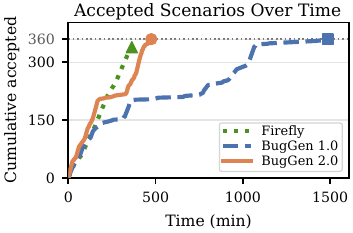}
    \vspace{-1mm}
    {\footnotesize\textbf{(c)} FabScalar cumulative acceptance (G4)}
\end{minipage}

\vspace{1mm}

\begin{minipage}[t]{0.32\textwidth}
    \centering
    \includegraphics[width=\linewidth]
        {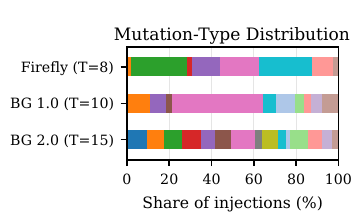}
    \vspace{-1mm}
    {\footnotesize\textbf{(d)} AES mutation distribution (G4)}
\end{minipage}
\hfill
\begin{minipage}[t]{0.32\textwidth}
    \centering
    \includegraphics[width=\linewidth]
        {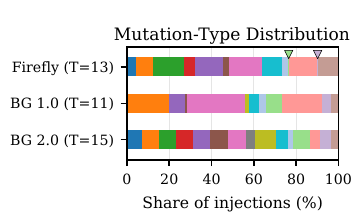}
    \vspace{-1mm}
    {\footnotesize\textbf{(e)} AES mutation distribution (D4F)}
\end{minipage}
\hfill
\begin{minipage}[t]{0.32\textwidth}
    \centering
    \includegraphics[width=\linewidth]
        {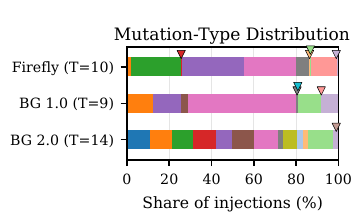}
    \vspace{-1mm}
    {\footnotesize\textbf{(f)} FabScalar mutation distribution (G4)}
\end{minipage}

\par\vspace{3.5mm}
\includegraphics[width=0.98\textwidth]{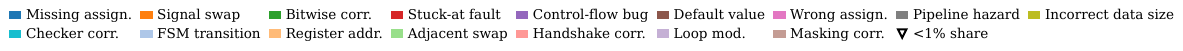}
\vspace{-1.0mm}

\caption{End-to-end convergence and mutation type representation with Gemma 4 and DeepSeek v4 Flash.
(a)--(c) Cumulative accepted bug scenarios over run time, with the dotted horizontal lines marking target quotas.
(d)--(f) Mutation type distributions in accepted scenarios.}
\label{fig:efficiency_diversity}
\end{figure*}

\begin{table}[!ht]
\caption{BugGen 2.0 on additional OpenTitan designs.}
\label{tab:bg2_scalability}
\centering
\footnotesize
\setlength{\tabcolsep}{3.2pt}
\renewcommand{\arraystretch}{1.00}
\begin{tabular}{@{}llccccc@{}}
\toprule
\textbf{Design} & \textbf{Model} & \textbf{A/Att.}
& \textbf{C} & \textbf{kTok/A} & \textbf{T} & \textbf{S} \\
\midrule
OTPCtrl & G4  & 120/124 & 67\% & 4.24 & 16 & 120 \\
        & OSS & 120/121 & 78\% & 4.89 & 16 & 120 \\
        & D4F & 120/123 & 74\% & 3.84 & 16 & 120 \\
\midrule
RomCtrl & G4  & 150/160 & 73\% & 4.17 & 14 & 165 \\
        & OSS & 150/156 & 71\% & 6.16 & 15 & 170 \\
        & D4F & 150/157 & 77\% & 3.75 & 15 & 170 \\
\midrule
USB     & G4  & 160/166 & 77\% & 4.23 & 14 & 173 \\
        & OSS & 160/169 & 80\% & 5.48 & 14 & 182 \\
        & D4F & 160/165 & 85\% & 3.74 & 15 & 180 \\
\bottomrule
\end{tabular}
\end{table}

The additional OpenTitan results show that this consistency extends beyond AES and FabScalar. All nine model and design combinations meet their quotas with low token cost and 14--16 represented mutation types across substantially different RTL blocks.

Fig.~\ref{fig:efficiency_diversity} complements these aggregate results with convergence trajectories and mutation type distributions. Firefly generates its full candidate set before simulation, so its final yield depends on the quality of that initial set. This works well on FabScalar, where 339 of 340 are accepted, but poorly on AES, where only 56 of 140 survive with both G4 and D4F. BugGen 1.0 instead retries unsuccessful scenarios, but because prior outcomes remain advisory, requeued scenarios can revisit similar targets and mutation choices, producing long retry loops with little progress. BugGen 2.0 avoids much of this behavior through explicit filtering and history-guided selection, leading to steadier convergence. The lower panels further show that this efficiency does not come from concentrating on a small subset of mutation types. BugGen 2.0 maintains both broad and highly consistent mutation type distributions across LLM backends, suggesting that the pipeline itself plays a larger role in preserving diversity than model-specific behavior.

\subsection{Line-Position Spread and Component Effects}

Table~\ref{tab:traditional_summary} gives a descriptive comparison of AES outcomes and line-position spread for the G4 runs in Table~\ref{tab:baseline_comparison}. For the LLM-based systems, D/F/U denotes accepted candidates, failed Verible or compiler validation, and compiled but sub-threshold candidates. For the commercial tool, the same letters denote detected, dropped, and undetected individual faults. The outcome categories, therefore, partition candidates within a system but are not equivalent across systems.

To measure how evenly accepted mutations are distributed across the RTL, each target file is divided into \(M=10\) equal line bins. These bins are used only for evaluation and are separate from the strata used during target selection. For LLM systems, \(p_i\) is the fraction of accepted mutation locations in bin \(i\). For the commercial tool, it is the fraction of reported static fault locations. We compute the normalized Shannon entropy as
\begin{equation}
\label{eq:spread}
\mathrm{Spread}
=
\frac{-\sum_{i=1}^{M} p_i \log_2 p_i}{\log_2 M}
\end{equation}

Scores are averaged across the seven AES target files. A score of 0 indicates complete concentration, while 1 indicates uniform placement. We assign 0 to target files with no accepted location. The metric measures source-line spread, not semantic or architectural coverage.

\begin{table}[!ht]
\caption{AES outcomes and line-position spread with Gemma 4.}
\label{tab:traditional_summary}
\centering
\small
\setlength{\tabcolsep}{5pt}
\renewcommand{\arraystretch}{1.00}
\begin{tabular}{@{}lcc@{}}
\toprule
\textbf{System} & \textbf{D/F/U} & \textbf{Spread} \\
\midrule
Commercial tool & 523/335/1   & 0.726 \\
Firefly         & 56/77/7     & 0.514 \\
BugGen 1.0      & 62/94/33    & 0.327 \\
BugGen 2.0      & 140/40/23   & \textbf{0.858} \\
\bottomrule
\end{tabular}
\end{table}

BugGen 2.0 has the highest line-position spread in Table~\ref{tab:traditional_summary}, while the partial BugGen 1.0 run leaves three target files with no accepted location. This pattern is consistent with stratified sampling favoring underexplored line ranges instead of repeatedly revisiting already explored ones. While differences in candidate conditioning, sample size, and scenario-versus-fault units prevent these statistics from serving as direct measures of overall fault quality, they provide a useful view of how broadly each method distributes accepted bugs across RTL locations.

Table~\ref{tab:ablation} reports independent ablations of pre-filtering, compile learning, and stratification on AES. The independent full run was conducted after the runs in Tables~\ref{tab:baseline_comparison} and~\ref{tab:traditional_summary}, so the results differ slightly. The run without pre-filtering terminated after the two-day wall-clock limit, at 117 accepted scenarios.

\begin{table}[!ht]
\caption{Component ablation on OpenTitan AES with Gemma 4.}
\label{tab:ablation}
\centering
\small
\setlength{\tabcolsep}{2.5pt}
\renewcommand{\arraystretch}{1.00}
\begin{tabular}{@{}lcccccc@{}}
\toprule
\textbf{Configuration} & \textbf{Acc.} & \textbf{C}
& \textbf{kTok/A} & \textbf{Sims} & \textbf{Types} & \textbf{Spread} \\
\midrule
Full BugGen 2.0          & 140 & 81\% & 4.4  & 159 & 15 & 0.862 \\
Without pre-filtering    & 117 & 91\% & 22.1 & 182 & 15 & 0.713 \\
Without compile learning & 140 & 78\% & 4.8  & 164 & 15 & 0.871 \\
Without stratification   & 140 & 83\% & 4.2  & 160 & 15 & 0.823 \\
\bottomrule
\end{tabular} 
\end{table}

The ablations separate the roles of the three mechanisms. Pre-filtering yields the greatest inference savings, as removing it causes no-op outputs and model requests to increase sharply, despite the higher validation rate among the candidates that survive generation. Compile learning reduces repeated invalid choices, while stratification determines how accepted mutations are distributed across the RTL. These results support using separate deterministic mechanisms for different sources of inefficiency rather than asking the LLM to manage them all.

\subsection{Variant Augmentation}

Table~\ref{tab:augmentation} evaluates how variant augmentation expands the dataset while reducing simulation effort. \emph{Budget} simulates only the subset selected in Section~\ref{sec:augmentation}, while \emph{All} and \emph{GT} evaluate every retained variant. On AES, \emph{Budget} and \emph{All} are independent runs with different generated variants, so they characterize the overall tradeoff but cannot validate the variants skipped by \emph{Budget} on the same set. On FabScalar, \emph{GT} continues the \emph{Budget} run by simulating exactly the variants it skipped, enabling direct validation of the estimate.

\begin{table}[!t]
\caption{Variant augmentation with Gemma 4.}
\label{tab:augmentation}
\centering
\footnotesize
\setlength{\tabcolsep}{2pt}
\renewcommand{\arraystretch}{0.98}
\begin{tabular}{@{}lcccccc@{}}
\toprule
\textbf{Metric}
& \multicolumn{3}{c}{\textbf{OpenTitan AES}}
& \multicolumn{3}{c}{\textbf{FabScalar}} \\
\cmidrule(lr){2-4}\cmidrule(lr){5-7}
& \textbf{Base} & \textbf{\textit{Budget}} & \textbf{\textit{All}}
& \textbf{Base} & \textbf{\textit{Budget}} & \textbf{\textit{GT}} \\
\midrule
\multicolumn{7}{@{}l}{\textbf{Data expansion}} \\
Variants retained    & 0 & 385 & 386 & 0 & 962 & 962 \\
Variants simulated   & --- & 140 & 386 & --- & 636 & 962 \\
Variant sims skipped & --- & 63.6\% & 0\% & --- & 33.9\% & 0\% \\
Variants accepted    & --- & 114 & 300 & --- & 600 & 916 \\
Variant acc. rate    & --- & 81.4\% & 77.7\% & --- & 94.3\% & 95.2\% \\
\midrule
\multicolumn{7}{@{}l}{\textbf{Execution}} \\
RTL instances        & 140 & 525 & 526 & 360 & 1,322 & 1,322 \\
Evaluation attempts  & 181 & 340 & 592 & 451 & 1,087 & 1,413 \\
Run time (in hours)  & --- & 12.3 & 27.3 & --- & 15.0 & 17.8 \\
\midrule
\multicolumn{7}{@{}l}{\textbf{Acceptance estimation}} \\
Estimated accepted  & --- & 298 & 258 & --- & 906 & 906 \\
Estimated error (\(E_v\)) & --- & --- & 14.0\% & --- & --- & 1.09\% \\
Assume all error (\(E_{\mathrm{all\text{-}acc}}\)) & --- & --- & 28.7\% & --- & --- & 5.0\% \\
\bottomrule
\end{tabular}
\end{table}

To quantify estimation accuracy, let \(A_v\) be the measured number of accepted variants, \(\hat A_v\) be the estimate, and \(|V|\) be the number of retained variants. The error \(E_v\) measures the relative difference between measured and estimated acceptance, while \(E_{\mathrm{all\text{-}acc}}\) measures the error from assuming that every retained variant is accepted.
{
\begin{equation}
E_v=\frac{|\hat A_v-A_v|}{A_v},\quad
E_{\mathrm{all\text{-}acc}}=\frac{|V|-A_v}{A_v}.
\end{equation}
}

AES illustrates the simulation tradeoff. \emph{Budget} skips 63.6\(\%\) of the retained variant simulations, but its skipped variants are not later evaluated, so its estimation error cannot be measured directly. The independent \emph{All} run instead checks the estimator on a separate fully simulated set. Its estimate predicts 258 accepted variants, while the full simulation accepts 300 of 386, yielding a 14.0\(\%\) error, compared to 28.7\(\%\) when every retained variant is assumed to succeed. \emph{Budget} and \emph{All} take 12.3 and 27.3 hours, respectively. While their different base runs and variant sets preclude a controlled speedup comparison, the measurements still illustrate the practical runtime benefit that motivates budgeted simulation.

FabScalar provides the same-set validation unavailable for AES \emph{Budget}. After simulating 636 of 962 variants, \emph{Budget} estimates 906 accepted variants across the full set. \emph{GT} then evaluates the remaining 326 variants and obtains 916 accepted variants, giving a 1.09\% estimation error while avoiding 33.9\% of the simulations during the budgeted run. This also improves on the 5.0\% error obtained by assuming that every retained variant is accepted.

The FabScalar estimate uses a cache of 437 prior outcomes from earlier base generation runs. Table~\ref{tab:augmentation} excludes the cost of collecting this history, so the cache has an upfront cost but can be reused across subsequent augmentation runs. The cache is frozen before \emph{GT} evaluates the skipped variants, preventing those outcomes from influencing the estimate and providing an independent test of how well previously collected evidence transfers to new variants.

\section{Conclusion}

In conclusion, BugGen 2.0 shows that scalable RTL bug synthesis benefits from separating structural decisions from semantic rewriting. Deterministic filtering, targeting, validation, and feedback reduce the number of repeated low-value decisions, while the LLM is reserved for localized edits. Across five designs and three LLMs, BugGen 2.0 reaches every quota with broad mutation type representation and consistent efficiency. In completed same-model comparisons on AES and FabScalar, it reduces token use by 6.8\(\times\)--15.4\(\times\) relative to BugGen 1.0 and 1.5\(\times\)--7.9\(\times\) relative to Firefly. Variant augmentation addresses the remaining simulation bottleneck by reusing locations from accepted scenarios, simulating only a subset of alternatives, and estimating the rest. Together, these mechanisms make ground-truth RTL bug generation more practical and scalable for debugging, evaluation, and testbench grading.

The current evaluation establishes compilation and test observability, but does not characterize how individual bugs propagate through a module or affect architectural behavior. Stronger pre-simulation predictors and propagation-aware analysis could further reduce simulation cost while providing richer ground truth for debugging and verification. Such analysis could distinguish bugs that merely trigger failures from those with deeper functional or architectural effects, extending BugGen 2.0 toward more informative bug characterization.

\ifanonymous
\else
\section*{Acknowledgment}
The complete acknowledgment and funding information will be included in the final version. Additional acknowledgment text may be added before submission.
\fi

\bibliographystyle{IEEEtran}
\bibliography{references}

\end{document}